\documentclass{aa}
\usepackage{txfonts}
\usepackage[psamsfonts]{euscript}
\usepackage{graphicx}
\usepackage{natbib}
\bibpunct{(}{)}{;}{a}{}{,}

\newcommand{\nc}{\newcommand}
\nc{\fracd}[2]{\displaystyle\frac{#1}{#2}}

\nc{\cm}   {\mathrm{cm}}
\nc{\DnuD} {\Delta\nu_\mathrm{D}}
\nc{\dFR}  {\nabla\cdot\mathrm{F}_\mathrm{R}}
\nc{\dv}   {\delta V}
\nc{\FLy}  {F_\mathrm{Ly}}
\nc{\FLa}  {F_{\mathrm{Ly}\alpha}}
\nc{\dFLa} {\nabla\cdot\mathrm{F}_{\mathrm{Ly}\alpha}}
\nc{\dfl}  {\nabla\cdot\mathrm{F}_\ell}
\nc{\Fnua} {F_{\nu1}}
\nc{\FR}   {F_\mathrm{R}}
\nc{\FRa}  {F_\mathrm{R1}}
\nc{\FRn}  {F_\mathrm{RN}}
\nc{\gcc}  {\mathrm{gm}\:\mathrm{cm}^{-3}}
\nc{\Ha}   {\mathrm{H}\alpha}
\nc{\Hb}   {\mathrm{H}\beta}
\nc{\Hg}   {\mathrm{H}\gamma}
\nc{\Inu}  {I_\nu(\mu)}
\nc{\La}   {\mathrm{L}\alpha}
\nc{\kms}  {\mathrm{km}\:\mathrm{s}^{-1}}
\nc{\mH}   {m_\mathrm{H}}
\nc{\nel}  {n_\mathrm{e}}
\nc{\Ta}   {T_a}
\nc{\Te}   {T_\mathrm{e}}
\nc{\Vs}   {V_\mathrm{s}}
\nc{\xH}   {x_\mathrm{H}}

\nc{\Cij}  {C_{ij}}
\nc{\Cik}  {C_{ik}}
\nc{\Rik}  {R_{ik}}
\nc{\Rij}  {R_{ij}}
\nc{\Yij}  {Y_{ij}}
\nc{\Yik}  {Y_{ik}}
\nc{\Zik}  {Z_{ik}}
\nc{\Zij}  {Z_{ij}}

\defcitealias{Fadeyev:Gillet:1998}{Paper~I}
\defcitealias{Fadeyev:Gillet:2000}{Paper~II}
\defcitealias{Fadeyev:Gillet:2001}{Paper~III}
\defcitealias{Fadeyev:2002}{Paper~IV}

\begin{document}

\title{The structure of radiative shock waves}
\subtitle{V. Hydrogen emission lines.}
\author{Yu.A. Fadeyev \inst{1} \and D. Gillet\inst{2}}
\institute{Institute for Astronomy of the Russian Academy of Sciences,
 Pyatnitskaya 48, 119017 Moscow, Russia
 \and
 Observatoire de Haute--Provence - CNRS, F-04870
 Saint--Michel l'Observatoire, France}
\offprints{D. Gillet, \email{gillet@obs-hp.fr}}
\date{Received / Accepted }

\abstract{
We considered the structure of steady--state plane--parallel radiative shock
waves propagating through the partially ionized hydrogen gas of temperature
$T_1 = 3000$~K and density $10^{-12}~\gcc\le\rho_1\le 10^{-9}~\gcc$.
The upstream Mach numbers range within $6\le M_1\le 14$.
In frequency intervals of hydrogen lines the radiation field was treated using
the transfer equation in the frame of the observer for the moving medium,
whereas the continuum radiation was calculated for the static medium.
Doppler shifts in Balmer emission lines of the radiation flux emerging
from the upstream boundary of the shock wave model were found to be roughly
one--third of the shock wave velocity: $-\dv\approx \frac{1}{3}U_1$.
The gas emitting the Balmer line radiation is located at the rear of the
shock wave in the hydrogen recombination zone where the gas flow velocity
in the frame of the observer is approximately one--half of the shock wave
velocity: $-V^*\approx\frac{1}{2} U_1$.
The ratio of the Doppler shift to the gas flow velocity of
$\dv/V^* \approx 0.7$ results both from the small optical thickness
of the shock wave in line frequencies and the anisotropy of the radiation
field typical for the slab geometry.
In the ambient gas with density of $\rho_1\ga 10^{-11}~\gcc$ the flux in
the $\Ha$ frequency interval reveals the double structure of the profile.
A weaker $\Hb$ profile doubling was found for $\rho_1\gtrsim 10^{-10}~\gcc$
and $U_1\lesssim 50~\kms$.
The unshifted redward component of the double profile is due to
photodeexcitation accompanying the rapid growth of collisional
ionization in the narrow layer in front of the discontinuous jump.
}

\titlerunning{The structure of radiative shock waves. V}
\authorrunning{Yu.A. Fadeyev \& D. Gillet}
\maketitle

\section{Introduction}

It is now a well--established fact that hydrogen emission lines are
a characteristic feature of radially pulsating stars of various types.
Strong hydrogen emission is observed in Mira type \citep{Joy:1947,Joy:1954},
W~Vir \citep{Abt:1954,Wallerstein:1959} and RV~Tau \citep{Preston:1962}
pulsating variables.
Moreover, in the spectra of Mira stars Balmer emission lines persist during
the major part of the pulsation period \citep{Joy:1947,Richter:Wood:2001}.
The intensity of the hydrogen emission seems to correlate with
the amplitude of the pulsation since RR~Lyr variables exhibit only
weak emission lines \citep{Preston:1964} and in classical Cepheids
(for example in $\beta$~Dor) hydrogen emission lines are scarcely
detected \citep{Hutchinson:1975}.
The hydrogen emission in the spectra of pulsating stars is thought
to be due to radiative cooling of the gas compressed by the shock wave
propagating through the stellar atmosphere in each pulsation cycle
\citep{Kraft:1959,Wallerstein:1959,Abt:Hardie:1960,Gorbatskii:1961}.

High resolution spectroscopy reveals the doubling of $\Ha$ and $\Hb$
emission profiles, whereas profiles of higher Balmer lines exhibit only
the asymmetry.
This feature is observed not only in Mira stars
\citep{Bidelman:Ratcliffe:1954,Fox:Wood:Dopita:1984,Gillet:1988,Woodsworth:1995}
but also in W~Vir and RV~Tau variables
\citep{Lebre:Gillet:1991,Lebre:Gillet:1992}.
\citet{Bidelman:Ratcliffe:1954} explained the $\Ha$ profile doubling observed
in the Mira type star T~Cen as an absorption reversal rather than a real
duplicity, the absorption resulting from the cool hydrogen gas above
the propagating shock wave.
\citet{Willson:1976} interpreted the double structure of the emission profiles
in terms of a spherically symmetric shock wave with a radial distance from the
center of the star appreciably larger than the radius of the photosphere.
According to this model the flux of the redshifted component
emerges from the back side of the shock wave moving outward from the observer.
\citet{Woodsworth:1995} modelled double $\Ha$ profiles as three emission
components of equal width, two of which are blended.
Thus, different phenomenological models demonstrate the ambiguity
existing so far in our understanding of the origin of the Balmer emission
lines produced by the shock waves in stellar atmospheres.

This paper is the fifth in our series on the structure of radiative shock waves.
In our previous Papers I--IV
\citep{Fadeyev:Gillet:1998,Fadeyev:Gillet:2000,Fadeyev:Gillet:2001,Fadeyev:2002}
we presented the method of computation and described results for
the structure of radiative shock waves propagating through partially
ionized  hydrogen gas with temperature and density typical for
atmospheres of pulsating late--type stars.
An advantage of this approach is that the gas dynamics, radiation field and
atomic level populations are considered self--consistently for the whole
shock wave model.
However, in all our previous papers the radiative transfer was treated in
a static medium approximation, so that we were unable to compare the
calculated monochromatic radiation flux with observed emission profiles.
Below we describe the shock wave models with Doppler shifts in the
line profiles computed from the transfer equation in the frame of the
observer.

\section{Method of computation}

As in our previous studies we consider a steady--state plane--parallel
shock wave propagating through a homogeneous medium consisting solely of
atomic hydrogen gas.
The shock wave model is represented by a flat finite slab moving together
with a discontinuous jump toward the observer at velocity $\Vs$
which is regarded as negative.
The unperturbed gas enters the upstream face of the slab with velocity
$U_1 = - \Vs$, and throughout this paper $U_1$ is referred as the shock
wave velocity.

The space coordinate $X$ is measured along the path of the gas element moving
through the slab and is within the range $X_1\le X\le X_N$, where $X_1$ and $X_N$
are the coordinates of the upstream and downstream faces of the slab,
respectively.
We set $X=0$ at the discontinuous jump where the gas flow velocity $U$ and
the gas density $\rho$ undergo an abrupt change.
At the upstream face of the slab the gas is assumed to be unperturbed,
whereas at the downstream slab boundary the electron temperature $\Te$ and
the hydrogen ionization degree $\xH$ are assumed to approach their
postshock asymptotic values.
In general, the space coordinates of both boundaries were determined from
trial calculations in such a way that the solution describing the structure
of the shock wave is independent of coordinates $X_1$ and $X_N$.

For the calculation of the shock wave structure we follow our previos studies
\citepalias{Fadeyev:Gillet:1998, Fadeyev:Gillet:2000} and solve
the initial value problem for the fluid dynamics and rate equations
treated as a system of ordinary differential equations,
while radiative transfer is treated as a two--point boundary value
problem.
To take into account the strong coupling between the gas flow and
the radiation field produced by this flow
the ordinary differential equations and the transfer
equation are solved iteratively.

The radiation transfer equation is solved with the improved Feautrier
method \citep{Rybicki:Hummer:1991}
and the slab is divided into $N=1.6\times 10^4$ cells with
$6000$ cells ahead of the discontinuous jump.
Space intervals $\Delta X_{j-1/2} = X_j - X_{j-1}$ increase in
a geometrical progression in both directions from
the cell interface $X_\EuScript{J} = 0$.
For the sake of convenience the variables at the cell centers nearest
to the discontinuous jump are denoted with superscripts minus and
plus, that is $X^- = X_{\EuScript{J}-1/2}$ and $X^+ = X_{\EuScript{J}+1/2}$.

The frequency range
$7\times 10^{13}\mbox{Hz}\le\nu\le 10^{16}\mbox{Hz}$
consists of a set of intervals with boundaries either at hydrogen ionization
thresholds or at limits of line frequency intervals.
In the present study we consider the hydrogen atom with $L=5$ bound levels and
the continuum, so that the full frequency range is represented by 15 intervals
for the continuum and by 10 intervals for lines.

The electron collision excitation rates were calculated with effective
collision strengths from \citet{Scholz:1990} for the $1\to 2$ transition,
from \citet{Callaway:1994} for the $1\to 3$ and $2\to 3$ transitions and
from \citet{Aggarwal:1991} for other bound--bound transitions.
The electron collision ionization rates were computed with polynomial fits
from \cite{Lennon:1986}.
The photoexcitation rates were calculated for Doppler profiles.

In general, the interaction of the radiation field with the moving medium
is described in terms of the comoving frame partial differential
transfer equation.
The method developed by \citet{Mihalas:1975} treats the comoving frame
transfer equation as a two--point boundary value problem with respect to
the spatial coordinate and as a Cauchy problem with respect to the frequency.
However our attempts to apply such an approach to the shock wave structure
failed due to numerical instability arising because of the existence
of the velocity jump.
The difficulties accompanying the solution of the comoving frame transfer
equation for nonmonotonic gas flow velocity fields were discussed earlier
by \citet{Mihalas:1980}.

In our study the ratio of the gas flow velocity to the speed of light is
confined to values of $U/c\lesssim 3\cdot 10^{-4}$, so that the only
significant effect of the moving medium is the Doppler shift in the
spectral lines.
The radiation flux within all line frequency intervals is less than
$\sim 10^{-3}$ of the total radiative
flux produced by the shock wave, so that we assume that without
significant loss of self--consistency the Doppler shifts can be regarded as
insignificant for the structure of the shock wave.
Such an assumption allowed us to calculate the shock wave model
iteratively with a static transfer equation.
When the iteration procedure converges, we solve the transfer equation
in line frequency intervals for the moving medium.
In support of this approach below we compare solutions of the transfer
equation for a moving and for a static medium.

In plane parallel geometry the transfer equation in the frame of
the observer is \citep{Mihalas:1978}
\begin{equation}
\label{rte}
\mu\frac{d\Inu}{dX} = \eta_\nu(\mu) - \chi_\nu(\mu)\Inu ,
\end{equation}
where
$\mu = \cos(\theta)$ is the directional cosine,
$\Inu$ is the specific intensity,
$\eta_\nu(\mu)$ and $\chi_\nu(\mu)$ are the emission and extinction
coefficients, respectively.
For brevity in Eq.~(\ref{rte}) we omit the functional dependencies on $X$.

The motion toward the observer is regarded as negative, whereas
the velocity in the frame of the shock $U$ is positive.
Velocities in both frames relate via
\begin{equation}
\begin{array}{llll}
V &=& V^- + U^- - U ,      &  (X < 0) ,\\[6pt]
V &=& V^+ + U^+ - U ,\quad &  (X > 0) ,
\end{array}
\end{equation}
where $V^- = U^- - U_1$ and $V^+ = U^+ - U_1$.
Thus, the postshock compression ratio expressed in terms
of the gas flow velocity in the frame of the observer is given by
\begin{equation}
\label{etaps}
\frac{\rho}{\rho_1} = \frac{\Vs}{\Vs - 2V^+ + V} .
\end{equation}
Though on the r.h.s. of Eq.~(\ref{etaps}) we have only the observer's frame
velocity, this expression is equivalent to Eq.~(8) given in
\citetalias{Fadeyev:Gillet:2001}
where the postshock compression ratio is expressed in terms of
the specific internal energy of the gas and the radiation flux.

The line opacity and line emissivity coefficients $\kappa_\nu^\ell(\mu)$
and $\eta_\nu^\ell(\mu)$ are calculated on the assumption of complete
redistribution, with the Doppler profile given by
\begin{equation}
\label{phiD}
\phi(\nu) = \frac{1}{\DnuD\sqrt{\pi}}
\exp\left[
- \left(\nu - \nu_0 + \fracd{\nu_0}{c}\mu V\right)^2/\DnuD^2
\right] ,
\end{equation}
where $\nu_0$ is the line--center frequency,
\begin{equation}
\DnuD = \frac{\nu_0}{c}\left(\frac{2k\Ta}{\mH}\right)^{1/2}
\end{equation}
is the Doppler width, and $\mH$ is the mass of hydrogen atom.

\section{Shock wave model}

\begin{figure}
\resizebox{\hsize}{!}{\includegraphics{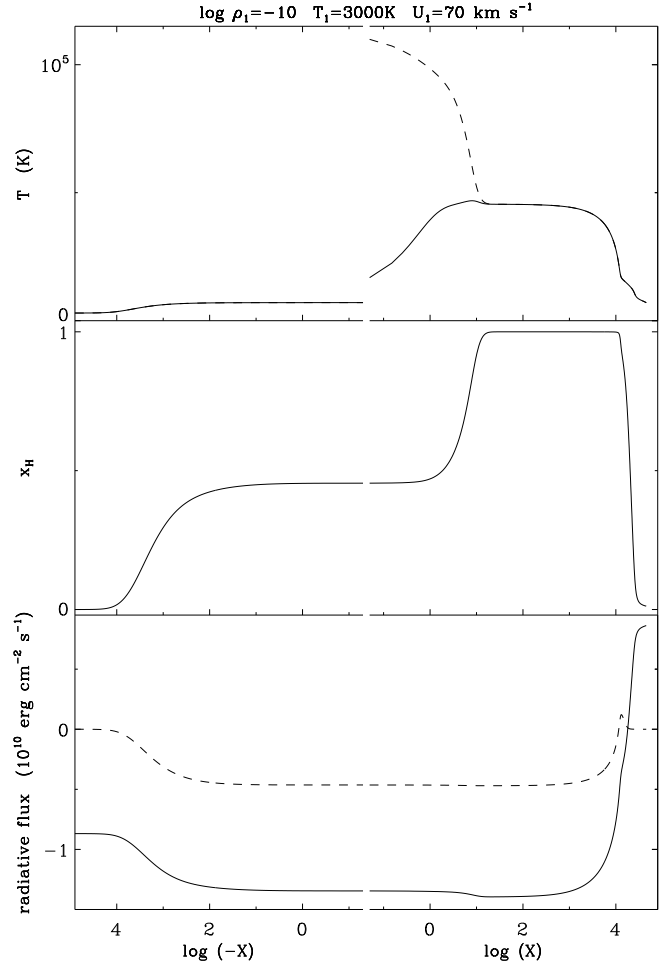}}
\caption{The structure of the radiative shock wave.
The left-- and right--hand sides of the plots
represent the preshock and postshock regions,
respectively.
Upper panel: the electron temperature $\Te$ (solid line) and the temperature
of neutral hydrogen atoms and hydrogen ions $\Ta$ (dashed line).
Middle panel: the hydrogen ionization degree $\xH$.
Lower panel: the total radiative flux $\FR$ (solid line) and
the Lyman continuum flux $\FLy$ (dashed line).
The space coordinate $X$ is in centimeters.}
\label{struct}
\end{figure}

The self--consistent solution of the equations of fluid dynamics,
radiative transfer and rate equations for atomic level populations describes
the structure of the shock wave and depends on three general parameters,
which are: the density $\rho_1$ and the temperature $T_1$ of the unperturbed
gas, and the velocity of the shock wave $U_1$ with respect to the ambient gas.
The structure of the radiative shock wave is shown in Fig.~\ref{struct}
where the following are plotted as a function of space coordinate $X$:
the electron temperature $\Te$ and the temperature of hydrogen atoms $\Ta$
(here we assume that neutral hydrogen atoms and hydrogen ions have the same
temperature), the hydrogen ionization degree $\xH$,
the total radiative flux $\FR$, and the Lyman continuum flux $\FLy$.
For the sake of convenience we use a logarithmic scale with respect to the
space coordinate, so that each variable is represented by two separate
plots for the preshock ($X<0$) and postshock ($X>0$) regions,
respectively.

In the partially ionized hydrogen gas the upstream boundary of the shock
wave is associated with ionization of hydrogen atoms ahead of the discontinuous
jump, whereas the downstream boundary is associated with the recombination zone.
Within the major part of the plots shown in the lower panel of
Fig.~\ref{struct} the total radiative flux is upstream (i.e. $\FR < 0$)
since the radiation is generated at the rear of the shock wave in the
hydrogen recombination zone.
The total radiative fluxes emerging from both faces of the slab
in opposite directions are nearly equal, that is, $-\FRa\approx\FRn$.

\begin{figure}
\resizebox{\hsize}{!}{\includegraphics{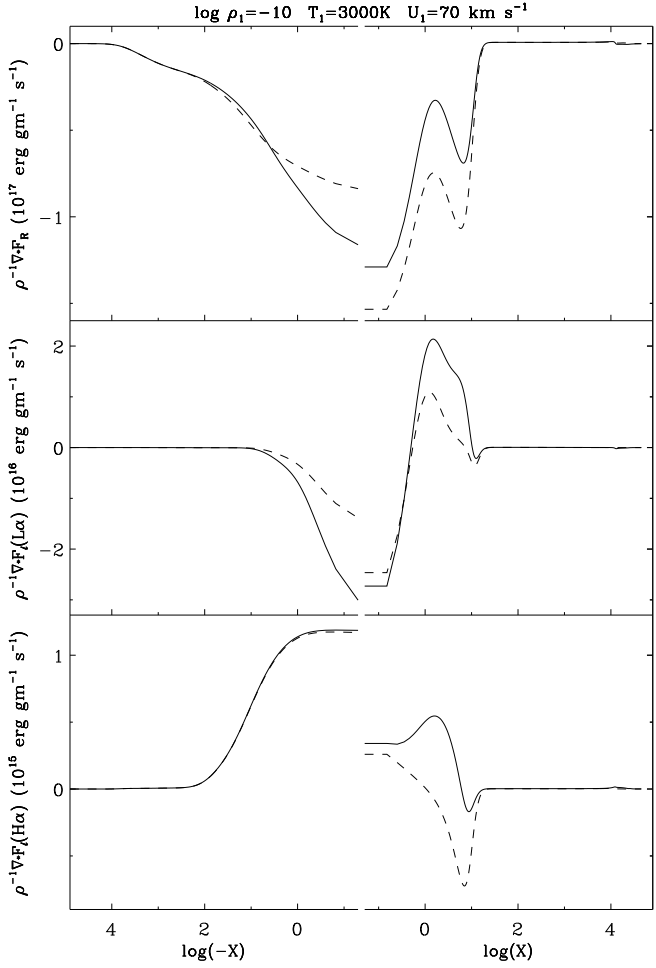}}
\caption{The divergence of the total radiative flux (upper panel),
of the $\La$ radiative flux (middle panel) and
of the $\Ha$ radiative flux (lower panel)
as a function of space coordinate $X$.
Solid and dashed lines show the solutions for the moving and for
the static medium, respectively.}
\label{divfr}
\end{figure}

The interaction of the gas flow with the radiative field due to absorption or
emission of radiation is described in terms of the divergence of the total
radiative flux
\begin{equation}
\label{dFR}
\dFR = 4\pi\int\limits_0^\infty
\left(\eta_\nu - \kappa_\nu J_\nu\right) d\nu ,
\end{equation}
where $J_\nu$ is the mean intensity.
In the upper panel of Fig.~\ref{divfr} we show the plots of $\dFR$
computed for the moving and for the static medium.
In the preshock region ($-10^4~\cm\lesssim X < 0~\cm$) the gas is radiatively
heated due to absorption of the Lyman continuum radiation.
Just ahead of the discontinuous jump ($-10^2~\cm\lesssim X < 0~\cm$)
radiative heating substantially increases because of absorption of radiation
in Lyman lines though at Balmer frequencies the gas emits more
energy than it absorbs.
The radiative cooling of the gas
due to emission in Balmer lines is less by an order of magnitude
than radiative heating due to absorption in Lyman lines.

Such a difference in behaviour of the radiation field in the $\La$ and $\Ha$
frequency intervals is clearly seen from the comparison of the plots of the
divergence of the line radiation flux
\begin{equation}
\label{dFl}
\dfl = 4\pi\int\limits_{\nu_0-\Delta\nu}^{\nu_0+\Delta\nu}
\left[\eta_\nu(\mu) - \kappa_\nu(\mu) J_\nu\right] d\nu
\end{equation}
shown in the middle and lower panels of Fig.~\ref{divfr}.
Here $\Delta\nu = 3.5\times 10^{-4}\nu_0$
is the half--width of the line frequency interval.
The opacity and emissivity coefficients are
$\kappa_\nu(\mu) = \kappa_\nu^\ell(\mu) + \kappa_\nu^c$ and
$\eta_\nu(\mu) = \eta_\nu^\ell(\mu) + \eta_\nu^c$, respectively,
where superscripts $\ell$ and $c$ denote the line and the continuum
variables.
Thus, expression (\ref{dFl}) defines the divergence of the radiative flux
for both line and continuum radiation.
It should be noted, however, that the contribution of the continuum radiation
can be neglected because the maximum values of the line opacity
and line emissivity coefficients dominate the continuum values
by as much as two orders of magnitude.

Most of the spectral line radiation is transported in Lyman series lines.
As is seen in the middle panel of Fig.~\ref{divfr}, Doppler shifts
lead to stronger radiative heating of the gas ahead of the discontinuous jump
and to stronger radiative cooling behind the discontinuous jump.
However the size of this zone is about four orders of magnitude less than
the length of the shock wave, so that the effects of Doppler shifts are
small enough and can be neglected.
For example, the change of the upstream $\La$ radiative flux due to
Doppler shifts does not exceed 10\%, whereas the $\La$ radiative flux
emerging from the upstream face of the slab is less than  $10^{-3}$
of the total radiative flux.
For the model shown in Figs.~\ref{struct} and \ref{divfr}
the optical thickness of the slab at $\La$ and $\Ha$ line--center frequencies
is $\tau_0(\La)\approx 6\times 10^6$ and $\tau_0(\Ha)\approx 40$,
respectively.

\begin{figure}
\resizebox{\hsize}{!}{\includegraphics{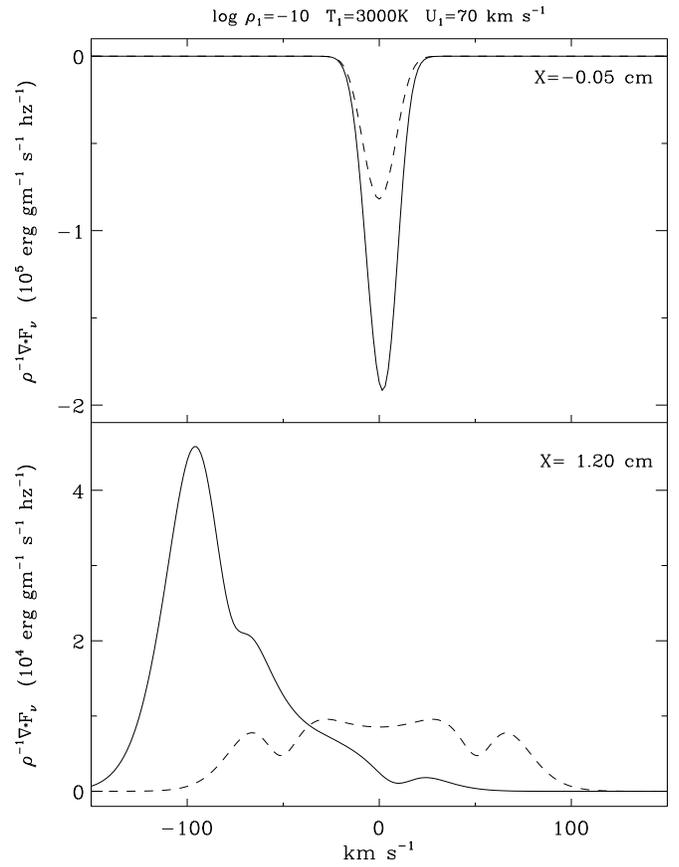}}
\caption{The divergence of the monochromatic radiative flux within
the $\La$ frequency interval just ahead of the discontinuous jump
(upper panel) and at the maximum of $\dfl(\La)$ (lower panel).
Solid and dashed lines show the solutions for the moving
and for the static medium, respectively.}
\label{dfla}
\end{figure}

Thus, the effects of Doppler shifts are most pronounced in the close
vicinity of the discontinuous jump.
In Fig.~\ref{dfla} we show the plots of the divergence of the
monochromatic radiative flux $\nabla\cdot\mathrm{F}_\nu(\La)$
as a function of frequency $\nu$ for the cell center $X^-$ just ahead the
discontinuous jump and for the maximum of $\dfl(\La)$ behind the
discontinuous jump (see the middle panel of Fig.~\ref{divfr}.

\section{Formation of line radiation}

Within the shock wave one can delineate three distinct zones with
substantial changes of level populations $n_i$ due to radiative
bound--bound transitions that lead to the appreciable emission of
spectral line radiation.
To compare the effect that different atomic transitions have on the
line radiation
in transition $i\to j$ we calculate the net photoexcitation rates
\begin{equation}
\Zij = n_i\Rij - n_j \frac{n_i^*}{n_j^*}\Rij^\dagger
\end{equation}
and the net collisional excitation rates
\begin{equation}
\Yij = \nel\Cij \left[n_i - n_j (n_i^*/n_j^*)\right] ,
\end{equation}
where $\Rij$ and $(n_i^*/n_j^*)\Rij^\dagger$ are the photoexcitation
and photodeexcitation rates, respectively,
$\Cij$ is the rate of collisional excitation and $n_i^*/n_j^*$ is
the thermodynamic equilibrium ratio given by the Saha--Boltzmann relation
for a nonequilibrium number density of electrons $\nel$.
We also estimate the net photoionization rates
\begin{equation}
\Zik = n_i\Rik - \nel\frac{n_i^*}{\nel^*}\Rik^\dagger ,
\end{equation}
where $\Rik$ and $(n_i^*/\nel^*)\Rik^\dagger$ are the
photoionization and photorecombination rates, respectively,
and the net collisional ionization rates
\begin{equation}
\Yik = n_i\nel\Cik\left(1 - \frac{\nel}{n_i}\frac{n_i^*}{\nel^*}\right) ,
\end{equation}
where $\Cik$ is the collisional ionization rate.

\subsection{The radiative precursor}

\begin{figure}
\resizebox{\hsize}{!}{\includegraphics{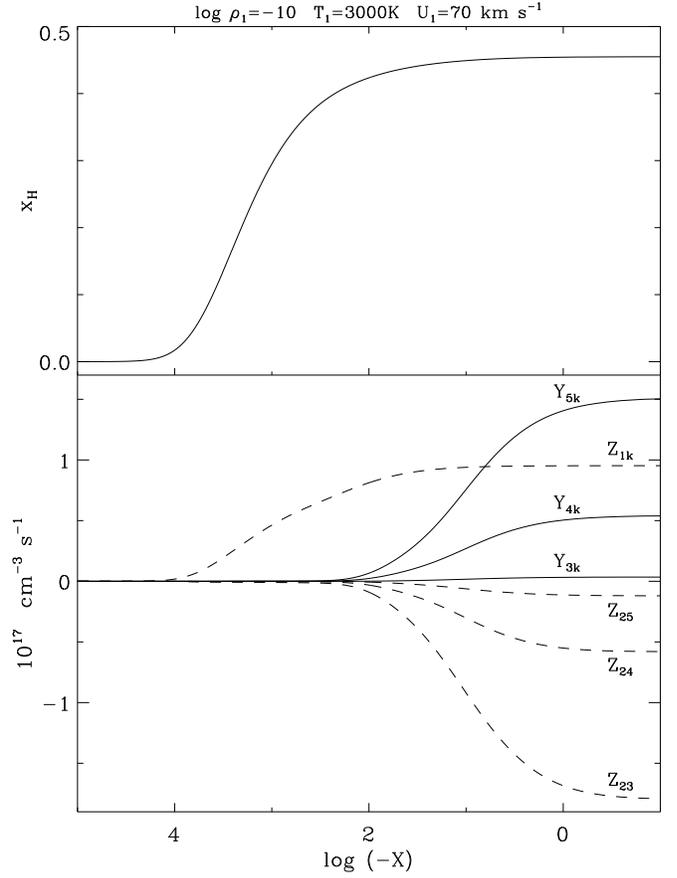}}
\caption{The zone of the radiative precursor.
Upper panel: the hydrogen ionization degree.
Lower panel: net collisional ionization rates (solid lines) and
net photoionization and photoexcitation rates (dashed lines).}
\label{rates1}
\end{figure}

As can be seen in Fig.~\ref{rates1} the only essential atomic transition
within the major part of the radiative precursor is photoionization from the
ground state which is due to the absorption of the Lyman continuum radiation.
However in the close vicinity of the discontinuous jump
($-10^2~\mbox{cm}\lesssim X\le 0$)
the rapid growth of collisional ionizations is accompanied by
photodeexcitation while the rates of collisional bound--bound transitions
remain negligible.
The effect of preshock photodeexcitation on the structure of the
shock wave is quite small; however, as we show below,
they lead to remarkable features in the $\Ha$ and $\Hb$ profiles
of the emergent radiation flux.

To estimate the contribution of preshock photodeexcitations $3\to 2$
on the emergent $\Ha$ radiation flux
\begin{equation}
\mathrm{F}_\ell(\Ha) =
\int\limits_{\nu_0(\Ha)-\Delta\nu}^{\nu_0(\Ha)+\Delta\nu} F_\nu d\nu
\end{equation}
we calculated the quantity
\begin{equation}
\label{dfha}
\EuScript{F}(\Ha) =
 - \int\limits_{X_1}^0 \nabla\cdot\mathrm{F}_\ell(\Ha) dX .
\end{equation}
The ratio $\EuScript{F}(\Ha)/F_\ell(\Ha)$ was found to decrease with
decreasing gas density $\rho_1$.
For example, in shock waves with velocity $U_1\approx 70~\kms$
the fraction of the preshock $\Ha$ emission in the
emergent radiation flux $\mathrm{F}_\ell(\Ha)$ is within the range
$10^{-2}\lesssim \EuScript{F}(\Ha)/F_\ell(\Ha)\lesssim 0.2$
for $10^{-12}~\gcc\le\rho_1\le 10^{-9}~\gcc$.

\subsection{The zone of hydrogen ionization}

\begin{figure}
\resizebox{\hsize}{!}{\includegraphics{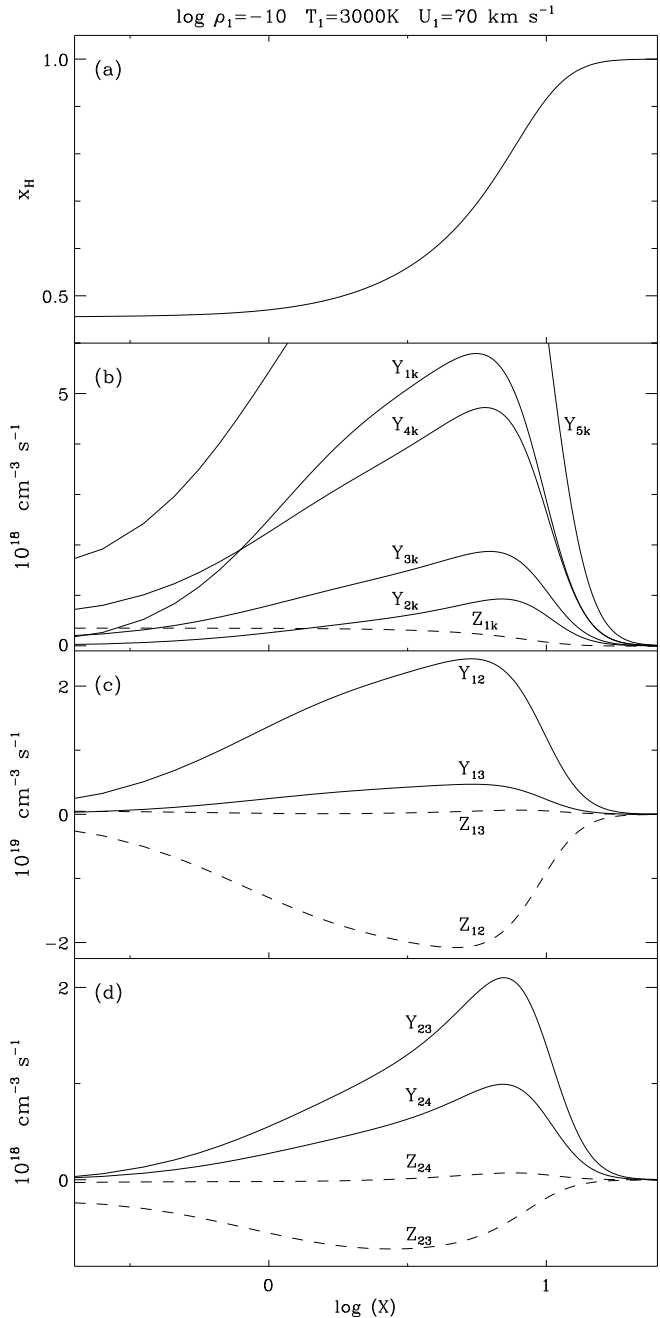}}
\caption{The zone of hydrogen ionization behind the discontinuous jump.
(a) The hydrogen ionization degree.
(b) Net collisional ionization rates (solid lines) and the
net photoionization rate $Z_{1k}$ (dashed line).
(c), (d) Net collisional excitation rates (solid lines) and net
photoexcitation rates (dashed lines).}
\label{rates2}
\end{figure}

Hydrogen ionization behind the discontinuous jump proceeds collisionally
since photoionization rates are much lower than the rapidly
increasing collisional ionization rates (see panel (b) in Fig.~\ref{rates2}).
The continuum radiation flux remains nearly constant so that any changes of
the total radiation flux behind the discontinuous jump result mainly
from $\La$ emission.
In panel (c) in Fig.~\ref{rates2} we show the net collisional
and photoexcitation rates for transitions $1-2$ and $1-3$.
Even though photodeexcitation $2\to 1$ does not exactly balance collisional
excitation $1\to 2$, it is responsible for strong emission of
$\La$ photons and leads to appreciable radiative cooling of the gas
seen in the middle panel in Fig.~\ref{divfr} as the maximum of $\dfl(\La)$.
Bound--bound transitions from the ground state to levels $j\ge 3$
are collisionally dominated and emission of radiation  due to
photodeexcitation is quite small.
Among excitations from levels $i > 2$ one should only note the collisional
transition $2\to 3$ accompanied by emission of $\Ha$ photons
(see panel (d) in Fig.~\ref{rates2}), which however is scarcely perceptible.

\subsection{The zone of hydrogen recombination}

The center of the radiation emitting zone we define as a layer where the
total radiative flux over the upstream hemisphere ($-1\le\mu\le 0$) equals
the total radiative flux over the downstream hemisphere ($0\le\mu\le 1$),
that is,
\begin{equation}
\label{fr0}
-\int\limits_0^\infty \int\limits_{-1}^0 \Inu \mu d\mu d\nu =
 \int\limits_0^\infty \int\limits_0^1 \Inu \mu d\mu d\nu .
\end{equation}
Thus, in the center of the radiation emitting zone we have $\FR=0$.
For the model shown in Fig.~\ref{struct} condition (\ref{fr0})
is fulfilled in the hydrogen recombination zone at $\xH=0.69$.
It should be noted, however, that for the monochromatic radiation flux $F_\nu$
the space coordinate $X(F_\nu=0)$ is frequency dependent.
For example, the radiative flux averaged over the Lyman continuum changes
its sign closer to the discontinuous jump in layers with $\xH\approx 0.99$.

\begin{figure}
\resizebox{\hsize}{!}{\includegraphics{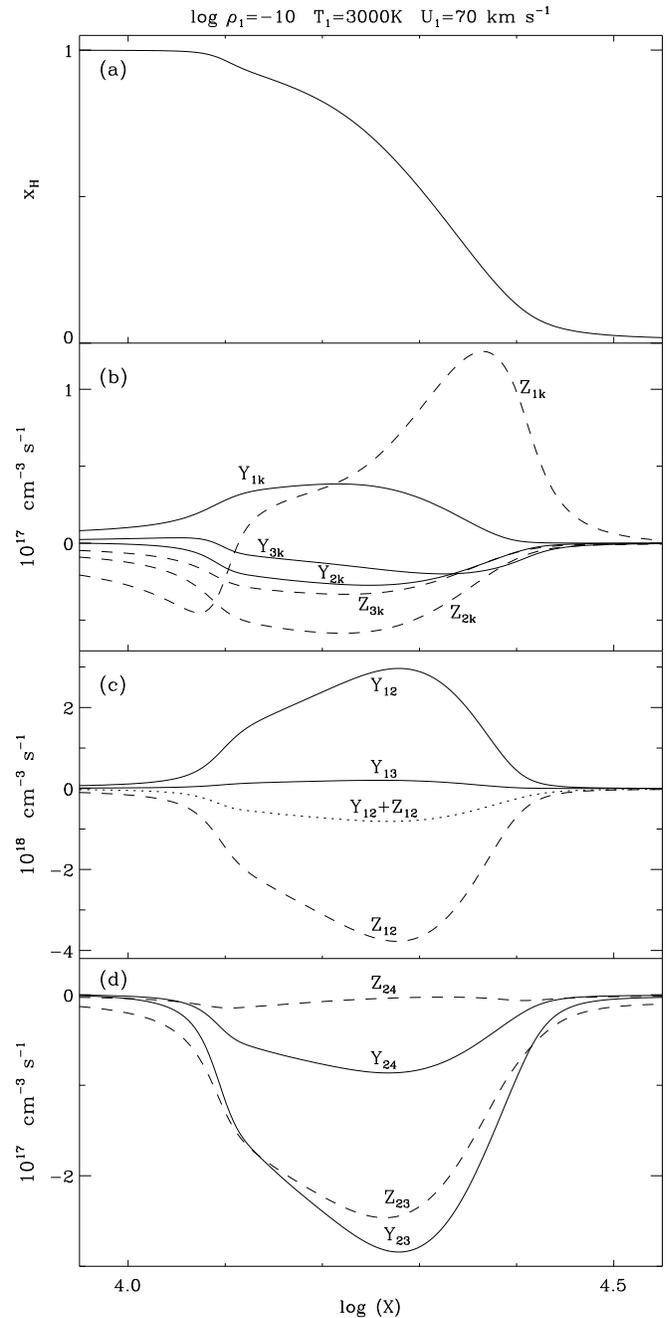}}
\caption{The zone of hydrogen recombination at the rear of the shock wave.
(a) The hydrogen ionization degree.
(b) Net collisional ionization rates (solid lines) and
net photoionization rates (dashed lines).
(c), (d) Net collisional excitation rates (solid lines) and net
photoexcitation rates (dashed lines).}
\label{rates3}
\end{figure}

As is seen in panel (b) in Fig.~\ref{rates3}, the initial hydrogen
recombination results from radiative downward transitions to the ground state.
However, at ionization degrees $\xH < 0.94$ recombination of hydrogen atoms
is due to downward bound--free transitions on levels $i\ge 2$, both radiative
and collisional, while atoms in the ground state undergo ionization.
Thus, the final relaxation of the postshock gas is appreciably slowed down
by ionization of the ground state hydrogen atoms.

In the recombination zone the ground--state hydrogen atoms undergo collisional
excitation, though radiative bound--bound transitions have the opposite
direction.
Within the recombination zone the most conspicuous transition is
excitation $1\to 2$ which is accompanied by photodeexcitation $2\to 1$
with resulting rate $Y_{12}+Z_{12} < 0$
(see panel (c) in Fig.~\ref{rates3}).
Though in the hydrogen recombination zone the absolute value of the net rate
$Z_{12}$ is about an order of magnitude less than in
the ionization zone (compare plots in panel (c) in
Fig.~\ref{rates3} with those in panel (c) in Fig.~\ref{rates2}),
most $\La$ photons of the shock wave are created in the
recombination zone because of its much larger length.
Collisional excitation $1\to j$ for $j\ge 3$ is not completely balanced by
photodeexcitation, and the resulting transition rates are positive, that is,
$Y_{1j}+Z_{1j} > 0$.

Populations of levels $i\ge 3$ undergo both collisional and
radiative deexcitation.
The net rates for transitions $2-3$ and $2-4$ are shown in panel (d) in
Fig.~\ref{rates3}.
Thus, the formation of the Balmer line radiation takes place at the rear of
the shock wave in the hydrogen recombination zone in layers around the
minimum of $Z_{2j} < 0$.
For the model shown in Fig.~\ref{rates3} the condition
$\mathrm{F}_\ell(\Ha) = 0$ corresponds to a hydrogen ionization degree
of $\xH\approx 0.87$, whereas for $\Hb$ and $\Hg$ frequency intervals
the condition of zero flux is fulfilled at $\xH\approx 0.74$.

\begin{figure*}
\resizebox{\hsize}{!}{\includegraphics{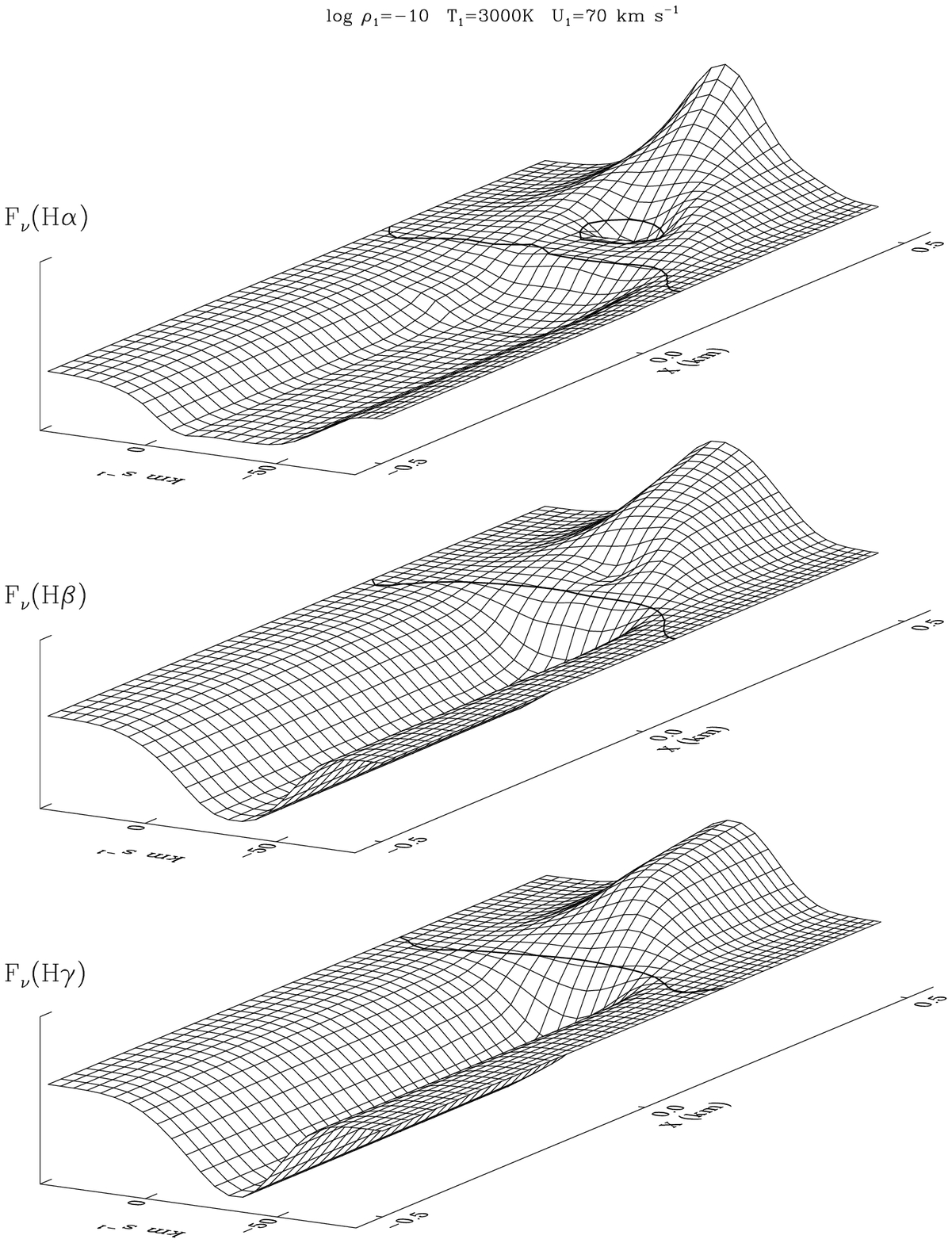}}
\caption{The monochromatic flux $F_\nu$ in the frequency intervals of
the $\Ha$, $\Hb$ and $\Hg$ lines.
Thick lines show the level of $F_\nu=0$.}
\label{balmer3d}
\end{figure*}

Transfer of the Balmer line radiation in the shock wave is illustrated in
Fig.~\ref{balmer3d} where the monochromatic radiation flux $F_\nu$
in $\Ha$, $\Hb$ and $\Hg$ frequency intervals is shown in three--dimensional
plots as a function of spatial coordinate $X$ and frequency $\nu$.
The plots are given on a linear scale with respect to $X$ and
for the sake of convenience we discarded the outermost preshock zones
where the flux $F_\nu$ is almost independent of $X$.
The frequency dependence of the flux is expressed in $\kms$
with negative values corresponding to the blue shifts.
The Balmer line--center flux is about two orders of magnitude
greater than the continuum flux, so that the dependence of the
continuum flux on the spatial coordinate $X$ is not seen in these plots.
According to notations adopted the upstream flux is negative.
The frequency dependence of the space coordinate of the zero
monochromatic flux $X(F_\nu=0)$ is shown by thick contour lines.

\section{The emergent flux at Balmer line frequencies}

\begin{figure*}
\resizebox{\hsize}{!}{\includegraphics{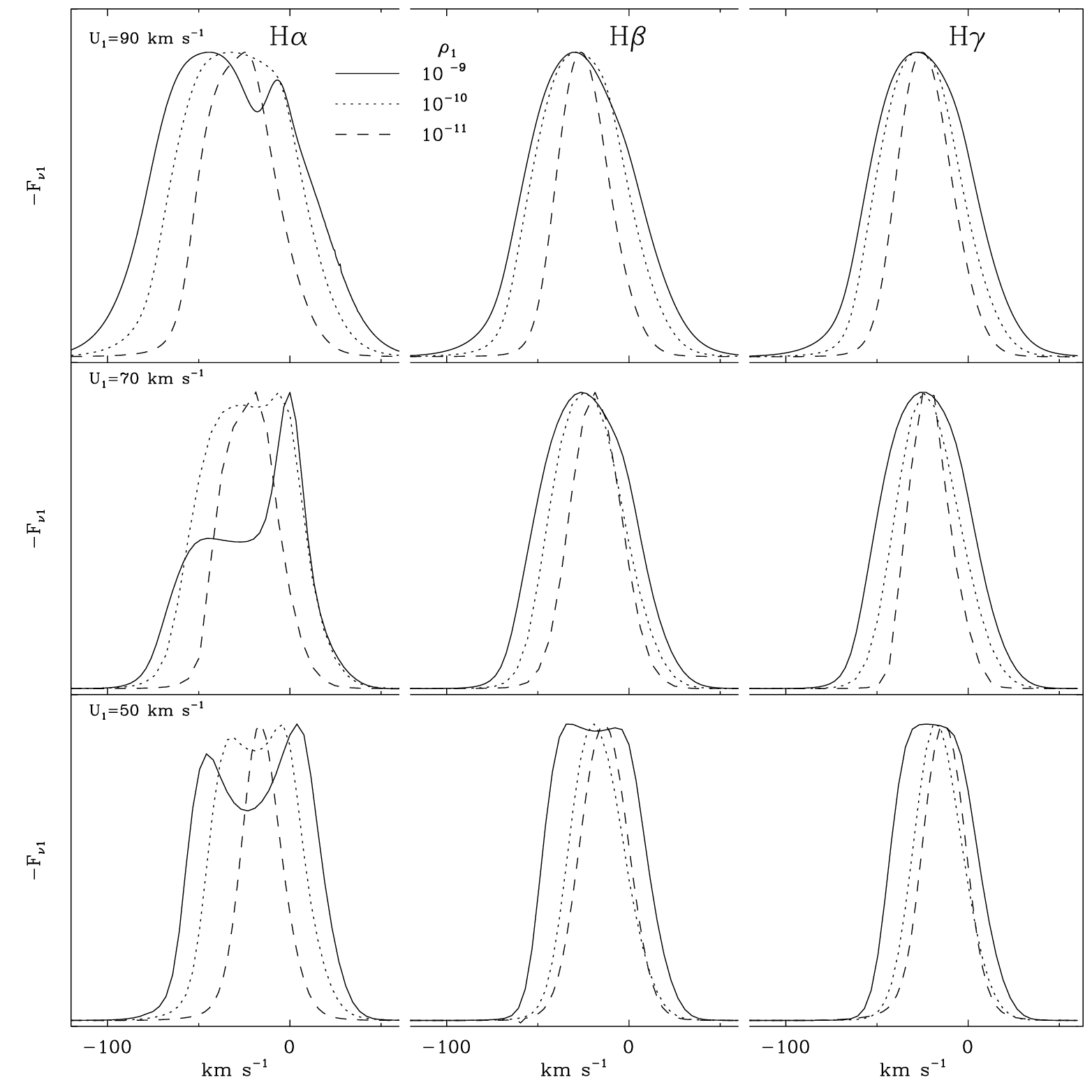}}
\caption{The normalized monochromatic flux at the upstream face of the slab
$\Fnua$ in the $\Ha$, $\Hb$ and $\Hg$ frequency intervals for shock wave models
with $\rho_1=10^{-11}$, $10^{-10}$, $10^{-9}~\gcc$, $T_1=3000$~K,
$U_1=50$, 70, $90~\kms$.}
\label{fnua}
\end{figure*}

Of most interest for a comparison with the observations is the spectral line
monochromatic radiation flux $\Fnua$ emerging from the upstream face
of the shock wave model.
In Fig.~\ref{fnua} we show the plots of $\Fnua$ as a function of
frequency $\nu$ for $\Ha$, $\Hb$ and $\Hg$ intervals.
For the sake of convenience the flux is normalized with respect to
the amplitude of the profile and is given with the opposite sign.
It should be noted that in order to treat correctly the angular
and frequency dependencies of the specific intensity $\Inu$, the line frequency
intervals considered in our calculations were several times of those shown
in Fig.~\ref{fnua}.

The most conspicuous feature seen in Fig.~\ref{fnua} is
an unshifted secondary maximum or a hump in the $\Ha$ and $\Hb$ profiles.
In the $\Ha$ frequency interval the hump is present at gas densities
of $\rho_1 > 10^{-11}~\gcc$ for all values of the shock wave velocity
considered in our study ($40~\kms\le U_1\le 90~\kms$),
while in the $\Hb$ frequency interval the weak hump appears only
at $\rho_1 > 10^{-10}~\gcc$ and $U_1 < 60~\kms$.
The hump becomes less prominent with decreasing gas density $\rho_1$ and
at low densities is revealed as an asymmetry of the redward branch of
the profile.

\begin{figure}
\resizebox{\hsize}{!}{\includegraphics{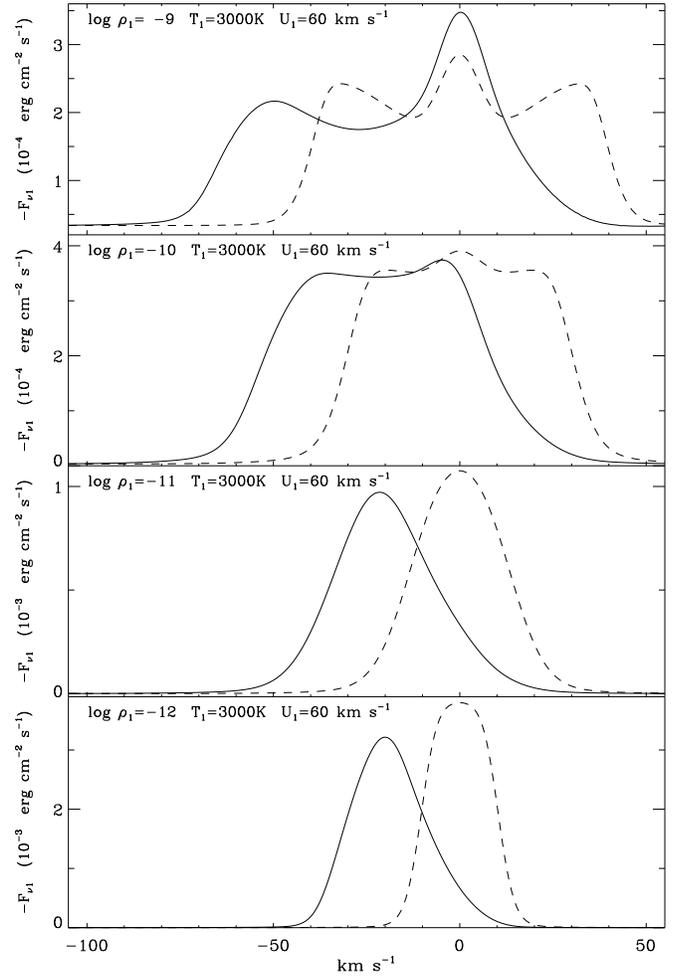}}
\caption{The monochromatic flux emerging from the upstream face of the
slab $\Fnua$ in the $\Ha$ frequency interval
for the moving (solid lines) and for the static (dashed lines) medium
for shock wave models with $10^{-12}~\gcc\le\rho_1\le 10^{-9}~\gcc$
and $U_1 = 60~\kms$.}
\label{halpha.u060}
\end{figure}

\begin{figure}
\resizebox{\hsize}{!}{\includegraphics{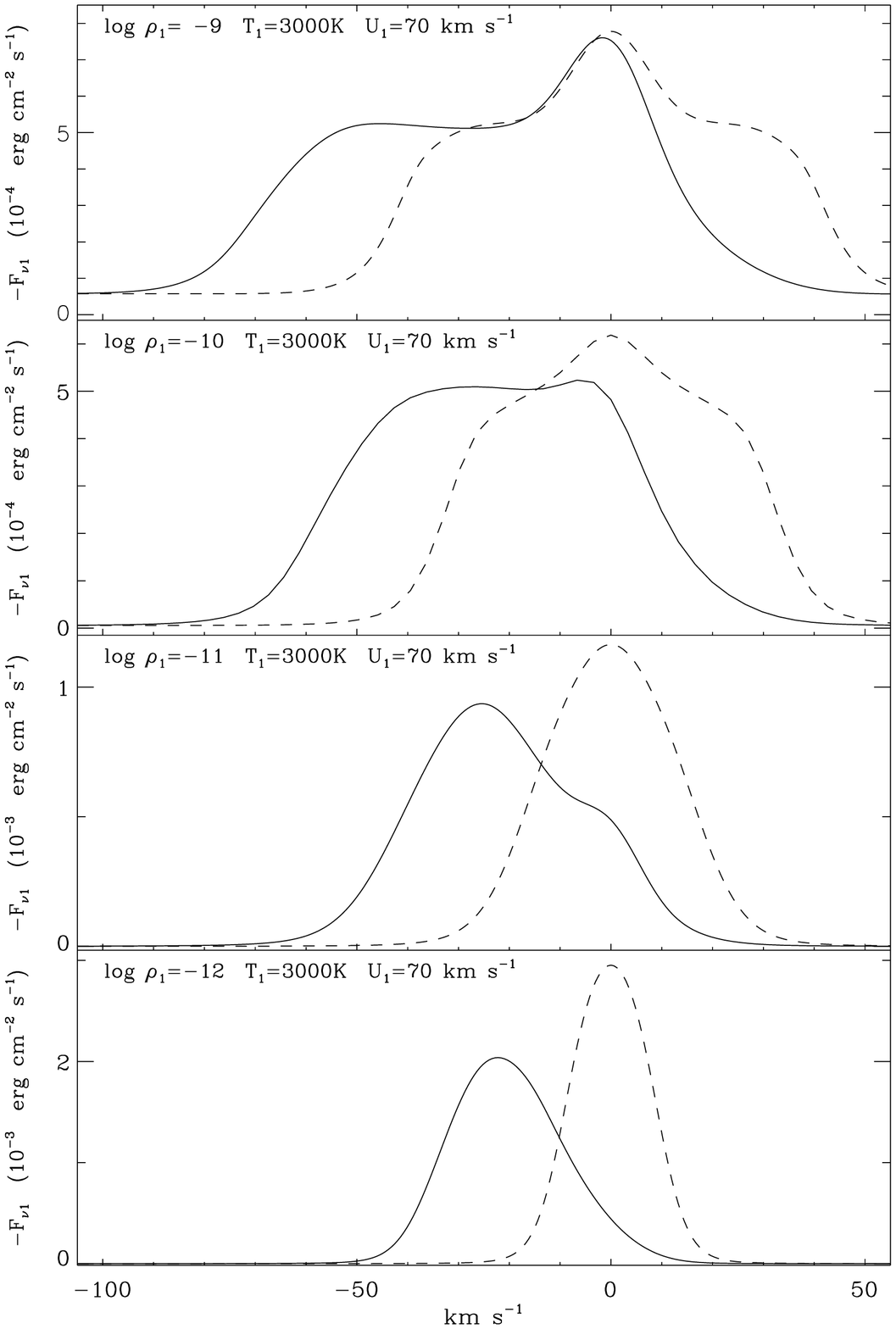}}
\caption{The same as in Fig.~\ref{halpha.u060}
for shock wave models with $U_1 = 70~\kms$.}
\label{halpha.u070}
\end{figure}

\begin{figure}
\resizebox{\hsize}{!}{\includegraphics{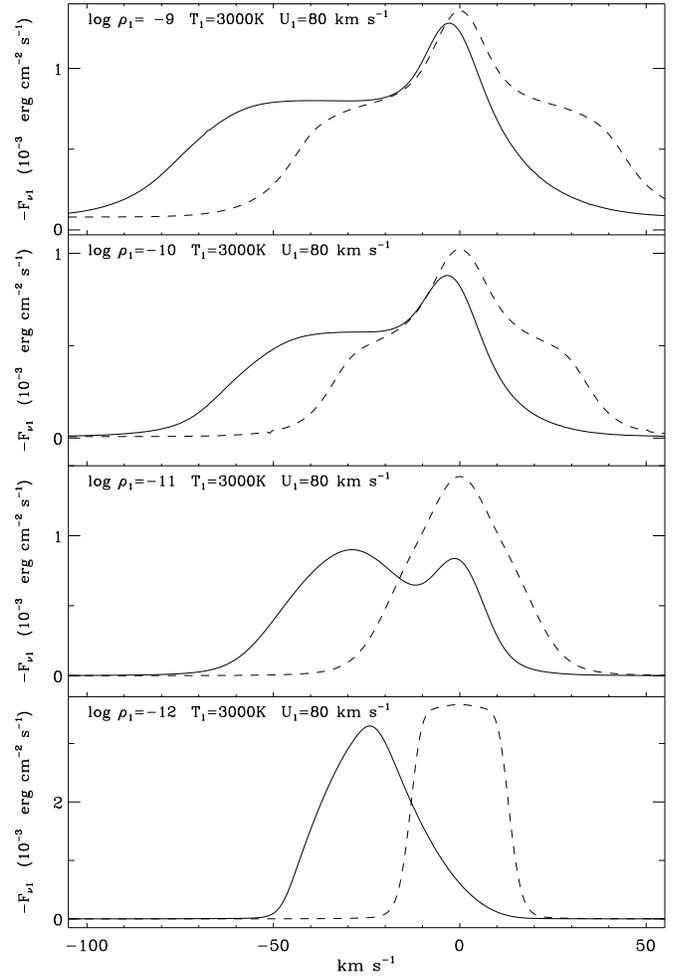}}
\caption{The same as in Fig.~\ref{halpha.u060}
for shock wave models with $U_1 = 80~\kms$.}
\label{halpha.u080}
\end{figure}

The hump in the $\Ha$ and $\Hb$ flux profiles results from the radiation
emitted by the narrow layer of the gas just ahead of the discontinuous jump.
The contribution from the preshock gas is clearly seen when comparing
the emergent flux profiles calculated for the moving and for the static medium.
As is seen in Figs.~\ref{halpha.u060} to \ref{halpha.u080},
the frequency dependence of the emergent $\Ha$ flux for the static medium
with $\rho_1\gtrsim 10^{-10}~\gcc$ can be represented as the superposition
of the narrow and the broad emission profiles with origin in the preshock
and postshock regions, respectively.
The preshock emission nearly disappears for the line--center
optical thickness of the shock wave of $\tau < 1$.

For $U_1 \ga 70~\kms$ the maximum of the preshock emission is blueshifted
by one or a few $\kms$.
This is due to the fact that the growth of the gas temperature
in the gas element approaching the discontinuous jump is accompanied by
an increase of the gas density $\rho$ and, therefore, by a decrease
of the gas flow velocity $U$.
Thus, the preshock gas ahead of the discontinuous jump is slightly accelerated
toward the observer and this effect is most conspicuous at higher densities
due to stronger absorption of the Lyman continuum radiation.
The Doppler shift of the redward emission component increases with
increasing shock wave velocity $U_1$ (see plots for $\log\rho_1 = -9$ in
Figs.~\ref{halpha.u060} to \ref{halpha.u080}) and
for shock wave velocities within the range $70~\kms\le U_1\le 90~\kms$
the gas flow velocity in the frame of the observer just ahead of
the discontinuous jump is $1~\kms\le -V^-\le 3~\kms$.

As was shown above, the preshock Balmer line emission zone is confined to
the narrow layer ahead of the discontinuous jump where photodeexcitation rates
rapidly increase while the gas element approaches the discontinuous jump
(see the lower panel in Fig.~\ref{rates1}).
The hump in the flux profile disappears when the contribution of preshock
radiative deexcitation becomes small in comparison with the line flux.
For $\Ha$ line the condition of disappearance of the redward emission feature
corresponds to $\EuScript{F(\Ha)}/F_\ell(\Ha) < 0.1$, where
$\EuScript{F(\Ha)}$ is given by Eq.~(\ref{dfha}).
The fraction of the preshock radiation in the line profile diminishes not only
with decreasing gas density $\rho_1$ but also with increasing
shock wave velocity $U_1$.
In the latter case this is due to the increasing width of the profile
accompanying the growth of the postshock temperature of heavy
particles $\Ta^+$.

To an accuracy of a few per cent the emergent line flux $F_\ell(\Ha)$ and the
FWHM of the line profile are the same for the moving and for the static medium,
respectively.
Because of the presence of the unshifted emission feature in the profiles
of the emergent $\Ha$ radiation flux any estimate of the Doppler shift
$\dv$ becomes ambiguous.
In the present paper we evaluated $\dv$ using the same procedure for all
profiles both with unshifted emission component and without it.
To this end we determined the Doppler shift of the profile $\dv$
as the frequency interval between
the profile center at its half height and the line--center frequency $\nu_0$.
Thus, values of $\dv$ might be underestimated at gas densities of
$\rho_1\gtrsim 10^{-10}~\gcc$ because of the presence of the unshifted
redward component.

The parameters of the shock wave models are summarised in Table~\ref{table}
where we give Doppler shifts $\dv$, FWHM line widths $w$ and optical depths
at line--center frequency $\tau$ for $\Ha$, $\Hb$ and $\Hg$
frequency intervals.
Gas flow velocities as well as Doppler shifts and line widths are
expressed in units of $\kms$.
In the first three columns of Table~\ref{table} are listed the unperturbed gas
density $\rho_1$, the shock wave velocity $U_1$ and the upstream Mach number
$M_1$.
The difference of the gas flow velocities in the frame of the observer
just ahead and
just behind the discontinuous jump $V^- - V^+$ gives the amplitude of
the discontinuity.
The columns labelled $V^*$ and $V_\infty$ give the velocity in the frame
of the observer for the $\Ha$ emitting layer
and the limiting gas flow velocity
corresponding to isothermal compression of the shocked gas.
The two last columns list the values of the compression ratio at the $\Ha$
emitting
layer $\eta^*=\rho^*/\rho_1$ and of the limiting isothermal compression ratio
\begin{equation}
\label{etainf}
\eta_\infty=\rho_\infty/\rho_1 = \gamma M_1^2 ,
\end{equation}
where $\gamma =5/3$ is the adiabatic exponent of the monoatomic gas.

\begin{table*}[t]
\begin{center}
\caption{Doppler shifts $\dv$, FWHM line widths $w$ and line--center
optical depths $\tau$ for $\Ha$, $\Hb$ and $\Hg$ frequency intervals.
The columns labelled $V^-$, $V^+$, $V^*$
give the gas flow velocity in the frame of the observer just ahead of
the discontinuous jump,
just behind the discontinuous jump and at the $\Ha$ emitting layer.
$V_\infty$ and $\eta_\infty$ are the isothermal limits of the gas flow
velocity in the frame of the observer and the compression ratio.
All models were computed for $T_1=3000$~K.}
\begin{tabular}{rrrrrrrrrrrrrrrrrrr}
\hline
  &    &    &  \multicolumn{3}{c}{$\Ha$} & \multicolumn{3}{c}{$\Hb$} & \multicolumn{3}{c}{$\Hg$} &   &    &    &    &    &  \\
$\rho_1$  & $U_1$ & $M_1$ &  $\delta V$ & $w$ & $\log\tau$ &  $\delta V$ & $w$ & $\log\tau$ &  $\delta V$ & $w$ & $\log\tau$ & $-V^-$ & $-V^+$ & $-V^*$ & $-V_\infty$ &$\eta^*$ & $\eta_\infty$\\
\hline
      -9 &    40 &  6.2 &   15 &   64 &  2.0 &   15 &   49 &  1.1 &   14 &   41 &  0.7 &    0 &   29 &   22 &   19 &   12 &   64 \\
         &    50 &  7.8 &   18 &   70 &  2.4 &   18 &   54 &  1.5 &   17 &   47 &  1.0 &    0 &   37 &   27 &   24 &   15 &  101 \\
         &    60 &  9.3 &   22 &   76 &  2.6 &   21 &   58 &  1.7 &   20 &   52 &  1.3 &    0 &   44 &   32 &   28 &   18 &  145 \\
         &    70 & 10.9 &   25 &   82 &  2.8 &   24 &   63 &  1.9 &   23 &   58 &  1.4 &    1 &   51 &   36 &   32 &   19 &  197 \\
         &    80 & 12.5 &   29 &   88 &  2.9 &   27 &   67 &  2.0 &   27 &   64 &  1.6 &    2 &   58 &   43 &   35 &   10 &  258 \\
         &    90 & 14.0 &   32 &   94 &  3.0 &   30 &   72 &  2.1 &   30 &   69 &  1.7 &    3 &   64 &   44 &   36 &    9 &  327 \\[3pt]
     -10 &    40 &  6.2 &   15 &   45 &  0.7 &   14 &   28 & -0.6 &   13 &   24 & -0.5 &    0 &   29 &   20 &   19 &   32 &   64 \\
         &    50 &  7.8 &   18 &   52 &  1.1 &   17 &   34 &  0.2 &   16 &   29 & -0.2 &    0 &   37 &   26 &   24 &   26 &  101 \\
         &    60 &  9.3 &   20 &   58 &  1.4 &   20 &   39 &  0.5 &   19 &   34 &  0.1 &    0 &   44 &   32 &   28 &   17 &  145 \\
         &    70 & 10.9 &   23 &   65 &  1.6 &   23 &   45 &  0.7 &   22 &   39 &  0.3 &    0 &   51 &   36 &   32 &   19 &  197 \\
         &    80 & 12.5 &   26 &   72 &  1.8 &   26 &   50 &  0.9 &   25 &   44 &  0.5 &    1 &   58 &   40 &   35 &   20 &  258 \\[3pt]
     -11 &    60 &  9.3 &   21 &   25 &  0.2 &   17 &   33 & -1.3 &   16 &   29 & -1.0 &    0 &   44 &   30 &   29 &   60 &  145 \\
         &    70 & 10.9 &   23 &   32 &  0.6 &   19 &   34 & -1.0 &   19 &   31 & -0.5 &    0 &   51 &   34 &   33 &   63 &  197 \\
         &    80 & 12.5 &   26 &   39 &  0.8 &   22 &   35 & -0.6 &   21 &   33 & -0.4 &    1 &   58 &   36 &   35 &      &  258 \\[3pt]
\hline
\end{tabular}
\label{table}
\end{center}
\end{table*}

Within the considered range of shock wave velocities $40~\kms\le U_1\le 90$
(i.e. for Mach numbers ranging within $6.2\le M_1\le 14$)
both Doppler shifts $\dv$ and FWHM line widths $w$ increase almost linearly
with increasing shock wave velocity $U_1$.
At the same time $\dv$ and $w$ decrease with decreasing density of the
unperturbed gas $\rho_1$.

\begin{figure}
\resizebox{\hsize}{!}{\includegraphics{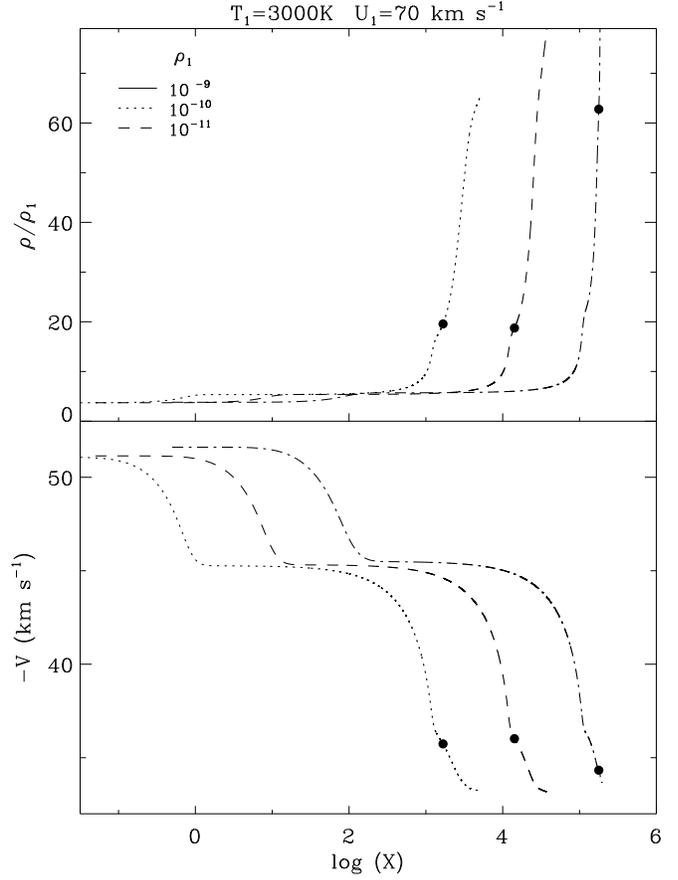}}
\caption{The postshock compression ratio (upper panel) and the postshock
gas flow velocity in the frame of the observer (lower panel) in shock waves with
$U_1=70~\kms$.
Solid lines: $\rho_1=10^{-9}~\gcc$; dotted lines: $\rho_1=10^{-10}~\gcc$,
dashed lines: $\rho_1=10^{-11}~\gcc$.
Filled circles indicate the layers of the formation of Balmer line radiation.}
\label{vobs}
\end{figure}

As is seen in Table ~\ref{table}, the Doppler shifts are roughly one third
of the shock wave velocity: $\dv\approx\frac{1}{3}U_1$.
There are two reasons for this.
First, most of the Balmer line radiation is emitted at the rear of
the shock wave where the gas undergoes strong compression and,
therefore, slows down appreciably.
In the upper panel in Fig.~\ref{vobs} we show the plots of the
postshock compression ratio $\rho/\rho_1$ for shock wave models
with $U_1=70~\kms$.
Filled circles in the curves indicate layers with $F_\ell(\Ha)=0$ where
the compression ratio is $\eta^* = \rho^*/\rho_1 \approx 20$ for
$10^{-10}~\gcc\le\rho_1\le10^{-9}~\gcc$ and
$\eta^* \approx 63$ for $\rho_1=10^{-11}~\gcc$.
Substituting these values of $\eta^*$ into Eq.~(\ref{etaps})
we find the gas flow velocity
in the layers where  Balmer line radiation is formed to be:
$-V^* = 36~\kms$ for
$10^{-10}~\gcc\le\rho_1\le10^{-9}~\gcc$ and
$-V^* = 34~\kms$ for $\rho_1=10^{-11}~\gcc$.
Thus, in the $\Ha$ emitting layer the gas flow velocity in the frame
of the observer is roughly one half of the shock wave velocity.
Large values of the compression ratio $\eta^* = \rho^*/\rho_1$ for
$\rho_1 < 10^{-10}~\gcc$ are due to the small optical thickness at the $\Ha$
central frequency and, therefore, more rapid radiative cooling of the gas.

A consequence of the rapid postshock radiative cooling is that the gas flow
velocity in the $\Ha$ emitting layer almost reaches its limiting value.
Indeed, the upper limit of the final postshock compression ratio is given by
the approximation of the isothermal shock (\ref{etainf}).
For low ionization degrees ($\xH\ll 1$) the sound speed is almost
independent of the gas density, so that for $T_1 = 3000$~K and
$U_1 = 70~\kms$ we have $\rho_\infty/\rho_1 = 197$.
Substituting this value of the compression ratio into Eq.~(\ref{etaps})
we obtain a lower limit of
the postshock velocity in the frame of the observer of $-V_\infty = 33~\kms$,
whereas the gas flow velocity is $-V^*\approx 34~\kms$.
Plots of the postshock gas flow velocity in the frame of the observer
are shown in the lower panel in Fig.~\ref{vobs}.
However, the ratio of the Doppler shift to the gas flow velocity is
$\dv/V^*\approx 0.7$, i.e. the low gas flow velocity in the hydrogen
recombination zone does not account for the small values of the
Doppler shifts.

\begin{figure}
\resizebox{\hsize}{!}{\includegraphics{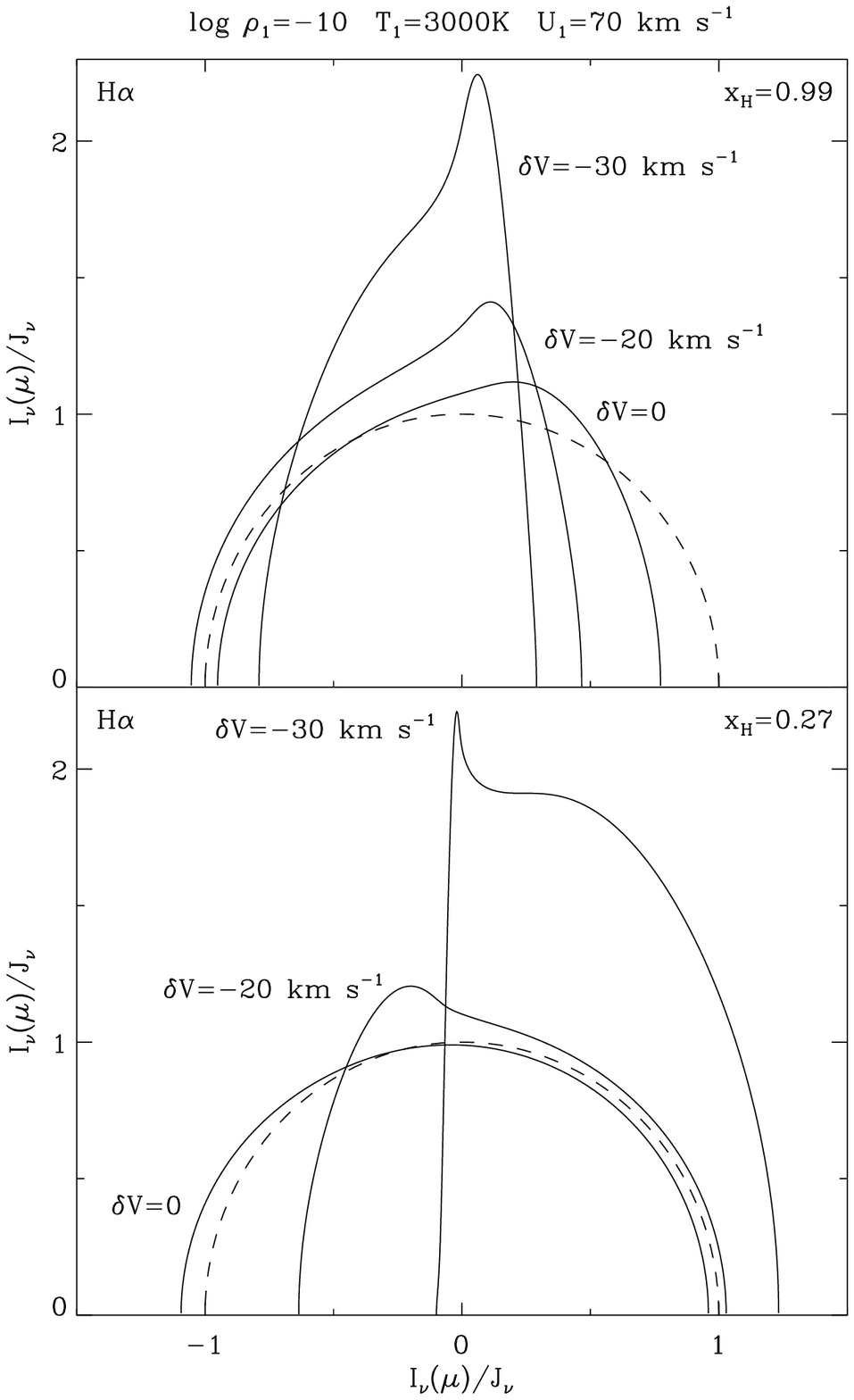}}
\caption{Polar diagrams of the normalized specific intensity $\Inu$
in the frequency interval of the $\Ha$ line in the hydrogen recombination zone.
Upper panel: layers with $F_\nu < 0$, lower panel: layers with $F_\nu > 0$.
Plots of $\Inu/J_\nu$ represent the static solution of the transfer equation.
The values of $\Delta V$ give the distance from the line--center frequency $\nu_0$
expressed in units of $\kms$.
Dashed line: the polar diagram of the isotropic radiation field.}
\label{imu}
\end{figure}

The second reason for the difference between $\dv$ and $V^*$ is that
the radiation field produced by the shock wave is characterized by an
anisotropy typical for slab geometry.
In Fig.~\ref{imu} we show the polar diagrams of the specific intensity
$\Inu$ for three points of the $\Ha$ frequency interval
with displacements from the line center of $\dv = 0$, 20 and $30~\kms$.
For the sake of simplicity the polar diagrams represent the solution of
the static transfer equation, and the specific intensity $\Inu$ is expressed
in units of the mean intensity $J_\nu$.
The radius vector connecting the origin with the point of the curve
is proportional to $\Inu$, whereas the angle between the
radius vector and the horizontal axis is the directional angle
$\theta = \cos^{-1}(\mu)$.
Polar diagrams in the upper and lower panels of Fig.~\ref{imu} correspond
to layers of the hydrogen recombination zone with $\FR < 0$ ($\xH=0.99$)
and $\FR > 0$ ($\xH=0.27$), respectively.
For the model represented in Fig.~\ref{imu} the condition $F(\Ha) = 0$
is fulfilled at $\xH=0.87$.

For the isotropic radiation field the polar diagram is the circle of unit
radius shown in Fig.~\ref{imu} by dashed lines.
Deviations from an isotropic radiation field are quite small near
the line center ($\dv\la 10~\kms$) but rapidly increase for
$\dv\ga 20~\kms$.
Thus, in calculating the monochromatic flux
\begin{equation}
F_\nu = \frac{1}{2}\int\limits_0^1 \Inu\mu d\mu
\end{equation}
we have a larger contribution of rays with
$\mu\ne\pm 1$ for $\dv\ga 20~\kms$ characterized by smaller
values of the projection of the velocity vector onto the normal.

\section{Discussion}

In our attempts to solve the
transfer equation for the shock wave structure in the co--moving frame
we encountered a severe difficulty because of the numerical
instability arising at the velocity discontinuity.
In the present study this difficulty could be circumvented because
the role of the radiation field in spectral lines is quite small in
comparison with that of the continuum.
According to our estimates the Doppler shifts in the hydrogen lines do not
affect perceptibly either the structure or the radiative losses of the shock wave.
In particular, the emergent flux integrated over the line frequency interval
was found to be the same within $\lesssim 1$\% for both the static and the
moving medium.
This allowed us to leave out the effects of the Doppler shifts from the global
iteration procedure and to solve the transfer equation in the frame of
the observer only in the final iteration.

The most remarkable result of our study is that the shock wave models
show the double emission structure in the $\Ha$ and $\Hb$
profiles of the emergent radiation flux which is well known from
high resolution high signal--to--noise ratio spectroscopy.
We showed that the redward emission feature results from the narrow
layer just ahead of the discontinuous jump within which the growth of
collisional ionization is accompanied by photodeexcitation onto the
second atomic level.
The contribution of preshock photodeexcitation decreases with decreasing
density of the ambient gas, therefore the double emission feature can be
considered as a tool for diagnostics of stellar atmospheres with
propagating shock waves.
Here one should bear in mind, however, that in the framework of our model
the unpertubed gas is at rest with respect to the observer, whereas in
pulsating stars the gas ahead of the shock wave falls down onto the star
with velocity in the range of one to a few dozen $\kms$.
Thus, the preshock emission feature in $\Ha$ and $\Hb$ profiles
should be observed as a redshifted component.

Another important conclusion is that the velocity inferred from
Doppler shifts of Balmer lines is roughly one--third of the shock wave
velocity: $\dv\approx \frac{1}{3} U_1$.
This is due to the fact that the gas layers emitting the Balmer line radiation
are located at the rear of the shock wave in the hydrogen recombination
zone where the velocity in the frame of the observer is roughly one half of
the shock wave velocity: $-V^*\approx\frac{1}{2}U_1$.
The ratio of the Doppler shift to the gas flow velocity of
$\dv/V^*\approx 0.7$ results from the small optical
thickness of the shock wave model and the anisotropy of the radiation field
produced by the shock wave.

$\Ha$ is the broadest emission line with FWHM comparable to the
velocity of the shock wave $U_1$.
This, as well as the significant contribution from the preshock
photodeexcitation zone make this emission line less appropriate for
inferring the velocity of the shock wave from observationally measured
Doppler shifts.
However the width of the emission profile is proportional to the temperature
of the hydrogen atoms behind the discontinuous jump $\Ta^+$ which is related
to the shock wave velocity $U_1$ via the Rankine--Hugoniot relations.
Thus, the width of the Balmer line is a function of two general quantities:
the shock wave velocity $U_1$ and the ambient gas density $\rho_1$.
In particular, the FWHM of $\Ha$ decreases by a factor of two
with decreasing gas density within $10^{-9}~\gcc\le\rho_1\le 10^{-12}~\gcc$.

Though our results are consistent with the observations,
there is a number of other parameters that determine the propagation of
the shock wave in the atmospheres of pulsating stars.
In particular, our model is confined to a flat finite slab
and therefore does not take into account the cool hydrogen gas
of the outer stellar atmosphere above the propagating shock wave.
Thus, we cannot exclude the role of absorption in the formation of the
double emission structure in Balmer lines.
The presence of such absorption in emission profiles observed
in RV~Tau and W~Vir stars was pointed out, for example, by
\cite{Lebre:Gillet:1991,Lebre:Gillet:1992}.

\begin{acknowledgements}
The work of YAF was done under the auspices of the Aix-Marseille I 
University, NATO and the Russian National Program ``Astronomy'' (item 1102).
\end{acknowledgements}

\end{document}